\documentclass[english,twocolumn]{webofc}
\begin{document}
\title{J/$\psi$ measurements in the STAR experiment}
% a short version of title for headings (if necessary)
\def\Title{J/$\psi$ measurements in the STAR experiment} 

\author{Barbara Trzeciak\inst{1} (for the STAR Collaboration)}
\def\Author{B. Trzeciak et al.} % a short version for author(s) names for headings

\institute{Faculty of Nuclear Sciences and Physical Engineering, Czech Technical University in Prague, Brehova 7, 115 19 Praha 1, Czech Republic }

\abstract{%
  In this paper, we present recent STAR J/$\psi$ results. J/$\psi$ nuclear modification factors ($R_{AA}$) in Au+Au collisions at $\sqrt{s_{NN}} =$ 200, 62.4 and 39 GeV and in U+U collisions at $\sqrt{s_{NN}} =$ 193 GeV are measured and compared to different theoretical calculations. We also report J/$\psi$ elliptic flow ($v_{2}$) results in Au+Au collisions at $\sqrt{s_{NN}} =$ 200 GeV and the first $\psi(2S)$ to J/$\psi$ ratio measurement in $p+p$ collisions at $\sqrt{s} =$ 500 GeV.
}
\maketitle
%

% -----------------------------------------------
\section{Introduction} 
\label{intro}

%Studies of production of quarkonium states in heavy-ion collisions can provide insight into the thermodynamic properties of the hot and dense medium~\cite{Mocsy:2007jz}. 
It was proposed that quarkonia are dissociated in the hot medium due to the Debye screening of the quark-antiquark potential and thus this ''melting'' can be a signature of Quark-Gluon Plasma (QGP) formation~\cite{Matsui:1986dk}. 
%Since different quarkonium states have different binding energies, they are expected to dissociate at different temperatures and can be treated as a QGP thermometer~\cite{Mocsy:2008eg}.
But there are other mechanisms that can alter quarkonium yields in heavy-ion collisions relative to $p+p$ collisions,  for example statistical recombination of heavy quark-antiquark pairs in the QGP or cold nuclear matter (CNM) effects. % - shadowing, final state nuclear absorption, initial-state parton energy loss.
Systematic measurements of the quarkonium production for different colliding systems, centralities and collision energies may help to understand the quarkonium production mechanisms in heavy-ion collisions as well as the medium properties.

% -----------------------------------------------
\vspace{-10pt}
\section{J/$\psi$ and $\psi(2S)$ measurements}
\label{sec:JpsiMeasurements}

%Charmonia at STAR are measured via the di-electron decay channel ($B_{ee} =$ 5.9\%) at mid-rapidity ($\vert y \vert < $ 1). 

\begin{figure}[ht]
		\centering
		\includegraphics[width=0.75\linewidth]{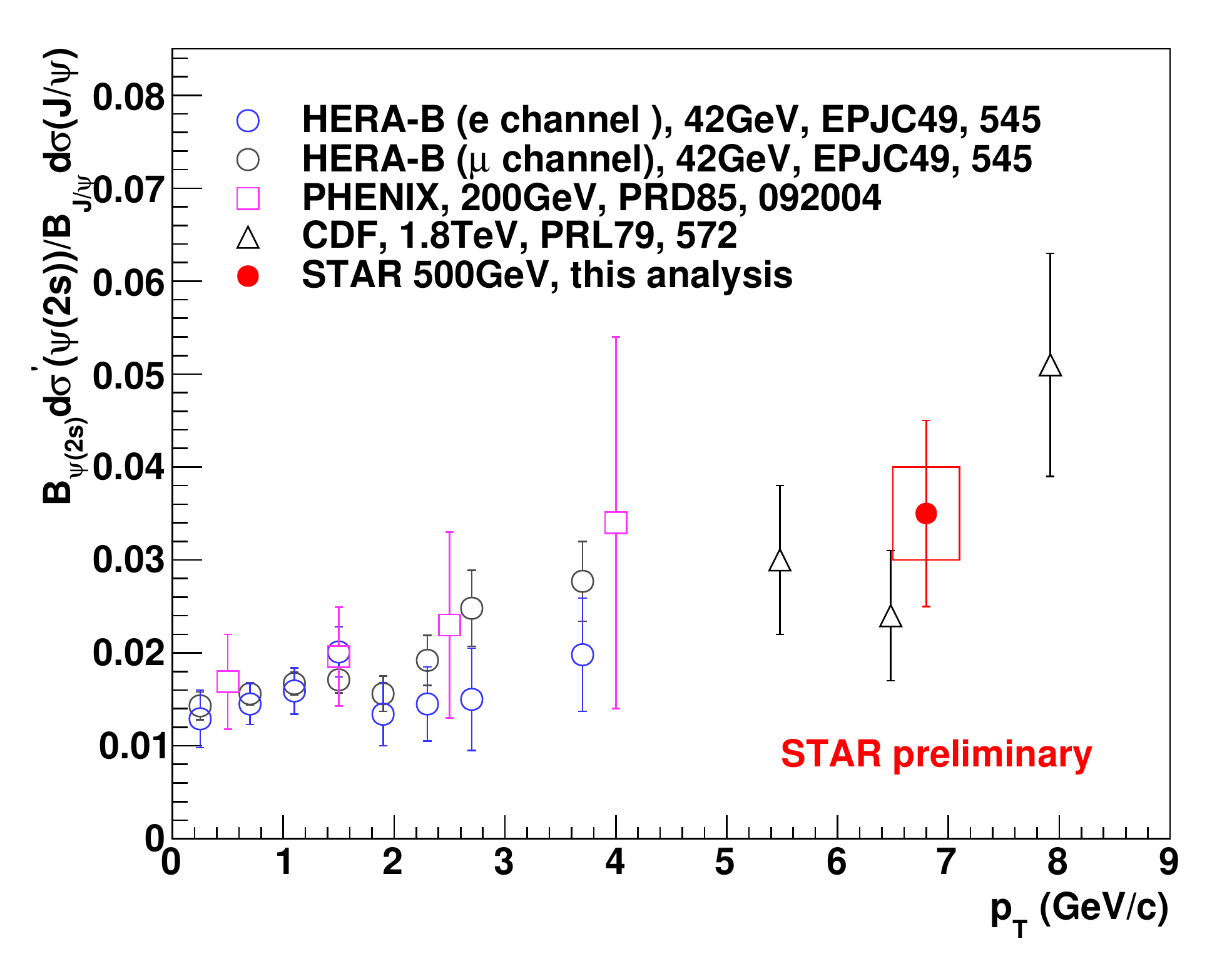}
		\caption{Ratio of $\psi(2S)$ to J/$\psi$ in $p+p$ collisions at $\sqrt{s} =$ 500 GeV from STAR (red circle) compared to results from other experiments at different energies.}
		\vspace{-10pt}
		\label{fig:psiRatio}
\end{figure}

\begin{figure}[ht]
		\centering
		\includegraphics[width=0.8\linewidth]{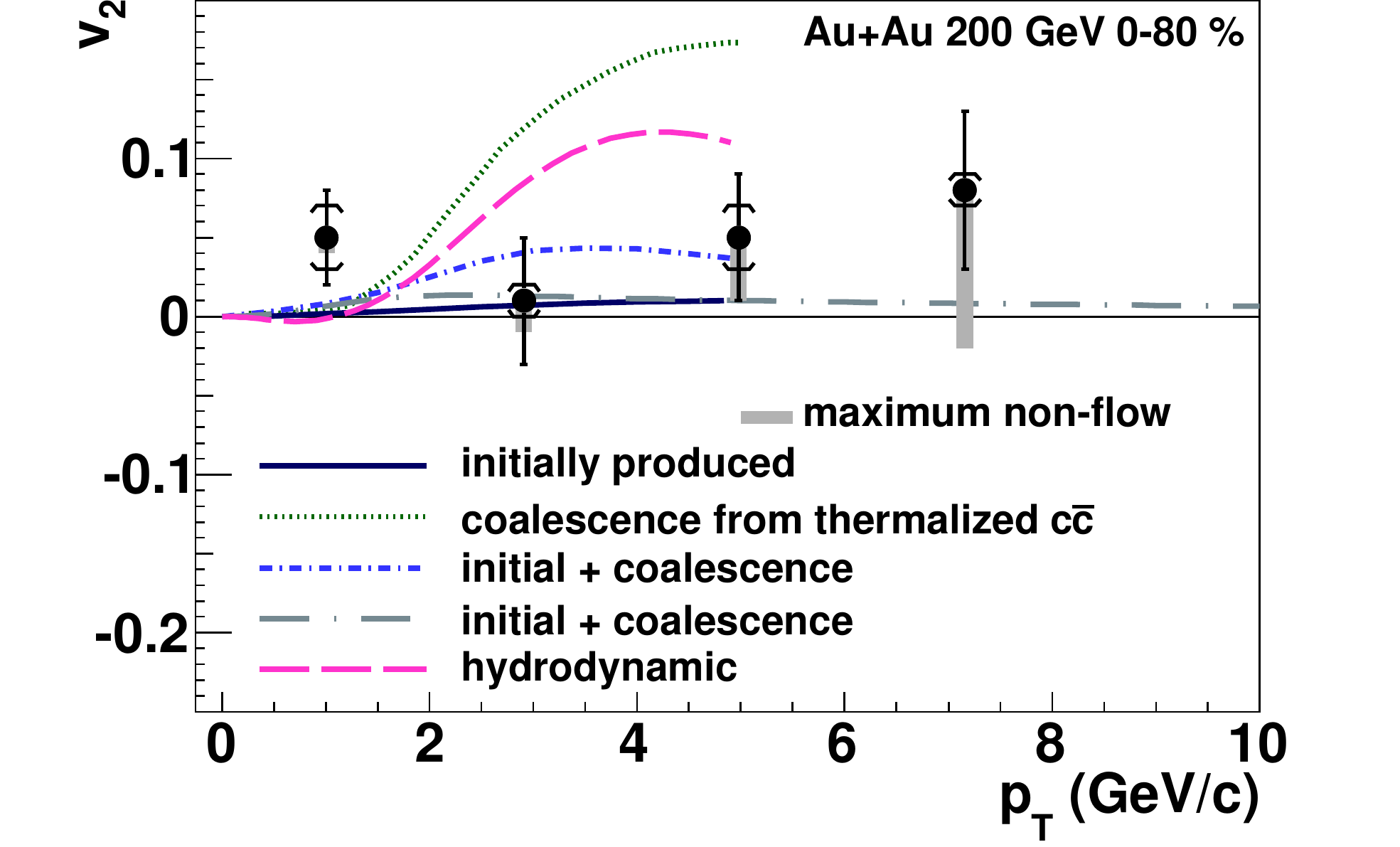}
		\caption{J/$\psi$ $v_{2}$ in Au+Au collisions at $\sqrt{s_{NN}}$ = 200 GeV at mid-rapidity in 0-80\% central events~\cite{Adamczyk:2012pw} with different model predictions (\cite{Yan:2006ve,Greco:2003vf,Zhao:2008vu,Liu:2009gx}). The gray boxes represent a non-flow estimation.}\vspace{-20pt}
		\label{fig:Jpsi_v2}
\end{figure}

\begin{figure}[ht]
		\centering
		\includegraphics[width=0.8\linewidth]{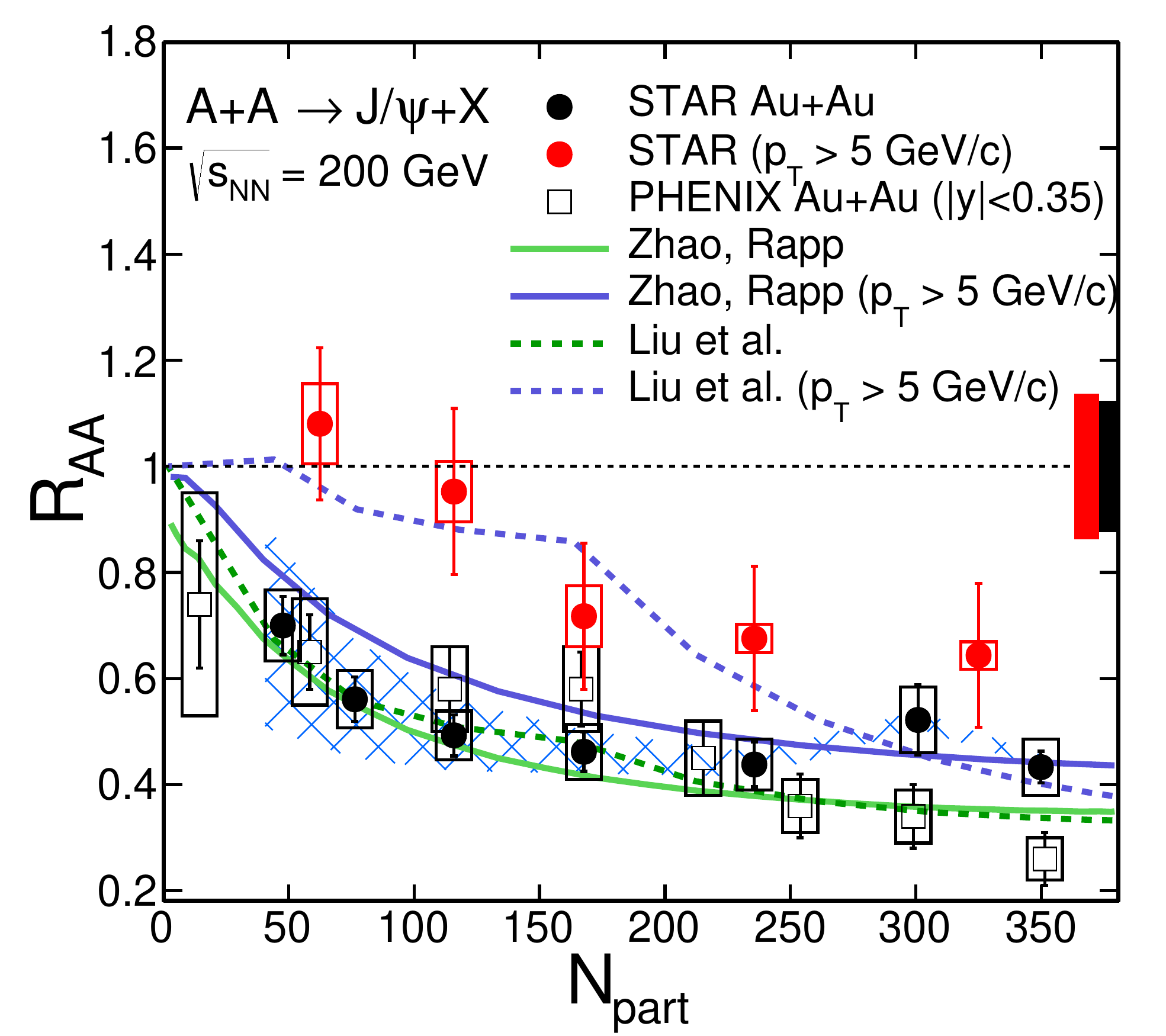}
		\caption{J/$\psi$ $R_{AA}$ as a function of $N_{part}$ in Au+Au collisions at $\sqrt{s_{NN}} =$ 200 GeV at mid-rapidity (\cite{Adamczyk:2012ey,Adamczyk:2013tvk}) with two model predictions (\cite{Zhao:2010nk,Liu:2009nb}). The low-$p_{T}$ ($<$ 5 GeV/$c$) result is shown as black full circles and the high-$p_{T}$ ($>$ 5 GeV/$c$) measurement as red full circles.}\vspace{-20pt}
		\label{fig:Jpsi_raa200}
\end{figure}

\begin{figure}[ht]
		\centering
		\includegraphics[width=0.8\linewidth]{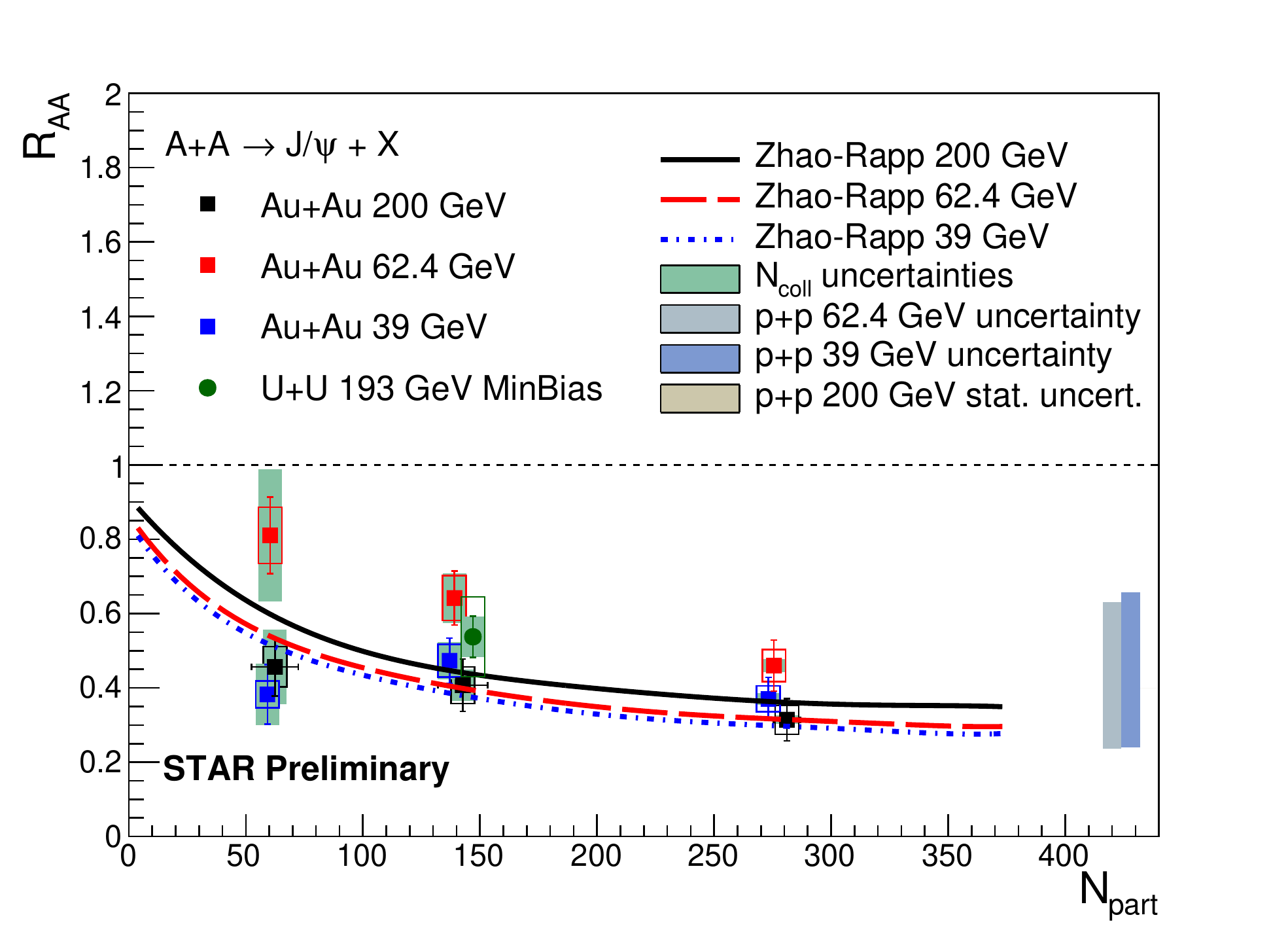}
		\caption{J/$\psi$ $R_{AA}$ as a function of $N_{part}$ in Au+Au collisions at $\sqrt{s_{NN}} =$ 200 (black), 62.4 (red) and 39 (blue) GeV at mid-rapidity with model predictions (\cite{Zhao:2010nk}). As the green circle the minimum bias U+U measurement at $\sqrt{s_{NN}} =$ 193 GeV is also presented\vspace{-20pt}.}
		\label{fig:Jpsi_raaBES}
\end{figure}

%J/$\psi$ production mechanism in elementary collisions is still not fully understood. 
STAR has measured J/$\psi$ $p_{T}$ spectra~\cite{Abelev:2009qaa,Adamczyk:2012ey} and polarization~\cite{Adamczyk:2013vjy} in $p+p$ collisions at $\sqrt{s} =$ 200 GeV via the dielectron decay channel ($B_{ee} =$ 5.9\%) at mid-rapidity ($\vert y \vert < $ 1). These results are compared to different model predictions to understand J/$\psi$ production mechanism in elementary collisions. In order to further test the charmonium production mechanism and constrain the feed-down contribution from the excited states to the inclusive J/$\psi$ production, the J/$\psi$ and $\psi(2S)$ signals were extracted in $p+p$ collisions at $\sqrt{s} =$ 500 GeV.
Figure~\ref{fig:psiRatio} shows $\psi(2S) / J/\psi$ ratio from STAR (red full circle) compared to measurements of other experiments at different colliding energies, in $p+p$ and $p+$A collisions. The STAR data point is consistent with the observed trend, and no collision energy dependence of the $\psi(2S)$ to J$/\psi$ ratio is seen with current precision.

In Au+Au collisions at $\sqrt{s_{NN}} =$ 200 GeV STAR has measured J/$\psi$ $p_{T}$ spectra for different centrality bins~\cite{Adamczyk:2013tvk,Adamczyk:2012ey}. It was found that at low $p_{T}$ ($\lesssim$ 2 GeV/$c$) the J/$\psi$ $p_{T}$ spectra are softer than the Tsallis Blast-Wave prediction, assuming that J/$\psi$ flows like lighter hadrons~\cite{Adamczyk:2013tvk}. This suggests that recombination may contribute to low-$p_{T}$ J/$\psi$ production. 
Measurement of J/$\psi$ $v_{2}$ may provide additional information about the J/$\psi$ production mechanisms. Figure \ref{fig:Jpsi_v2} shows J/$\psi$ $v_{2}$ measured in STAR in Au+Au collisions at $\sqrt{s_{NN}}$ = 200 GeV \cite{Adamczyk:2012pw}. At $p_{T} >$ 2 GeV/$c$ $v_{2}$ is consistent with zero. Compared to different model predictions \cite{Yan:2006ve,Greco:2003vf,Zhao:2008vu,Liu:2009gx}, data disfavor the scenario that J/$\psi$ with $p_{T} >$ 2 GeV/$c$ are dominantly produced by recombination (coalescence) from thermalized $c\bar{c}$ pairs.
Figure \ref{fig:Jpsi_raa200} shows J/$\psi$ $R_{AA}$ as a function of the number of participant nucleons ($N_{part}$) in Au+Au collisions at $\sqrt{s_{NN}}$ = 200 GeV, separately for low- ($<$ 5 GeV/$c$)~\cite{Adamczyk:2013tvk} and high-$p_{T}$ ($>$ 5 GeV/$c$)~\cite{Adamczyk:2012ey} regions. Suppression increases with collision centrality and the $R_{AA}$ at high $p_{T}$ is systematic higher than the low-$p_{T}$ one. The strong suppression of high-$p_{T}$ J/$\psi$ observed in central collisions (0-30\%) indicates color screening or other QGP effects -- at $p_{T} >$ 5 GeV/$c$ J/$\psi$ are expected to be less affected by the recombination and CNM effects. The $R_{AA}$ results are compared with two models, Zhao and Rapp \cite{Zhao:2010nk} and Liu {\it{et al.}} \cite{Liu:2009nb}. Both models take into account direct J/$\psi$ production with the color screening effect and J/$\psi$ produced via the recombination of $c$ and $\bar{c}$ quarks. The Zhao and Rapp model also includes the J/$\psi$ formation time effect and the B-hadron feed-down contribution. At low $p_{T}$ both predictions are in agreement with the data, while the high-$p_{T}$ result is well described by the Liu {\it{et al.}} model and the model of Zhao and Rapp underpredicts the measured $R_{AA}$.
%When changing energies of colliding heavy ions one expects different interplay between direct J/$\psi$ production (with color screening), CNM and recombination effects. 

Low-$p_{T}$ J/$\psi$ $R_{AA}$ measurements in Au+Au collisions at various colliding energies: $\sqrt{s_{NN}} =$ 200 (black), 62.4 (red) and 39 (blue) GeV are shown in Fig.~\ref{fig:Jpsi_raaBES}. Within the uncertainties, a similar level of suppression is observed for all three energies, which can be described by the  model predictions of Zhao and Rapp \cite{Zhao:2010nk}. However, it should be noted that due to lack of precise $p+p$ measurements at 62.4 and 39 GeV Color Evaporation Model calculations \cite{Nelson:2012bc} are used as baselines, which introduce large uncertainties.
Figure~\ref{fig:Jpsi_raaBES} also shows the Minimum Bias $R_{AA}$ measurement in U+U collisions at $\sqrt{s_{NN}} =$ 193 GeV as a full circle. In U+U collisions one can reach up to 20\% higher energy density compared to Au+Au collisions in the same centrality bin~\cite{Kikola:2011zz}. No difference in suppression compared to other measurements presented in Fig.~\ref{fig:Jpsi_raaBES} is observed.

% -----------------------------------------------
\vspace{-10pt}
\section{Summary}
\label{sec:summary}

In summary, significant suppression of low $p_{T}$ J/$\psi$ is seen in Au+Au collisions at various colliding energies: $\sqrt{s_{NN}}$ = 200, 62.4 and 39 GeV, and in U+U collisions at $\sqrt{s_{NN}} =$ 193 GeV. No strong energy dependence of the J/$\psi$ suppression in Au+Au is observed. Also, high-$p_{T}$ J/$\psi$ in Au+Au collisions at $\sqrt{s_{NN}}$ = 200 GeV  are strongly suppressed in central collisions, which suggests the QGP formation. $\psi(2S)$ to J/$\psi$ ratio was measured for the first time in $p+p$ collisions at $\sqrt{s} =$ 500 GeV.  When compared to results from other experiments, no collision energy dependence of the ratio is seen.
%New STAR upgrades will allow to perform more precise quarkonium measurements in next years.

% -----------------------------------------------
\vspace{-10pt}
\section*{Acknowledgements}

This publication was supported by the European social fund within the framework of realizing the project ,,Support of inter-sectoral mobility and quality enhancement of research teams at Czech Technical University in Prague'', CZ.1.07/2.3.00/30.0034. 

\vspace{-10pt}
\bibliography{/home/barbara/Work/Bibliography_all}

\begin{thebibliography}{14}

\bibitem{Matsui:1986dk}
T.~Matsui, H.~Satz, Phys.Lett. \textbf{B178}, 416 (1986)

\bibitem{Adamczyk:2012pw}
L.~Adamczyk et~al. (STAR Collaboration), Phys.Rev.Lett. \textbf{111}, 052301
  (2013), \texttt{1212.3304}

\bibitem{Yan:2006ve}
L.~Yan, P.~Zhuang, N.~Xu, Phys.Rev.Lett. \textbf{97}, 232301 (2006),
  \texttt{nucl-th/0608010}

\bibitem{Greco:2003vf}
V.~Greco, C.~Ko, R.~Rapp, Phys.Lett. \textbf{B595}, 202 (2004),
  \texttt{nucl-th/0312100}

\bibitem{Zhao:2008vu}
X.~Zhao, R.~Rapp (2008), \texttt{0806.1239}

\bibitem{Liu:2009gx}
Y.~Liu, N.~Xu, P.~Zhuang, Nucl.Phys. \textbf{A834}, 317C (2010),
  \texttt{0910.0959}

\bibitem{Adamczyk:2012ey}
L.~Adamczyk et~al. (STAR Collaboration), Phys. Lett. \textbf{B 722}, 55 (2013),
  \texttt{1208.2736}

\bibitem{Adamczyk:2013tvk}
L.~Adamczyk et~al. (STAR Collaboration), Phys.Rev. \textbf{C90}, 024906 (2014),
  \texttt{1310.3563}

\bibitem{Zhao:2010nk}
X.~Zhao, R.~Rapp, Phys.Rev. \textbf{C82}, 064905 (2010), \texttt{1008.5328}

\bibitem{Liu:2009nb}
Y.p. Liu, Z.~Qu, N.~Xu, P.f. Zhuang, Phys.Lett. \textbf{B678}, 72 (2009),
  \texttt{0901.2757}

\bibitem{Abelev:2009qaa}
B.~Abelev et~al. (STAR Collaboration), Phys. Rev. \textbf{C 80}, 041902 (2009),
  \texttt{0904.0439}

\bibitem{Adamczyk:2013vjy}
L.~Adamczyk et~al. (STAR Collaboration), accepted by Phys.Lett.B  (2014),
  \texttt{1311.1621}

\bibitem{Nelson:2012bc}
R.~Nelson, R.~Vogt, A.~Frawley, Phys.Rev. \textbf{C87}, 014908 (2013),
  \texttt{1210.4610}

\bibitem{Kikola:2011zz}
D.~Kikola, G.~Odyniec, R.~Vogt, Phys.Rev. \textbf{C84}, 054907 (2011),
  \texttt{1111.4693}

\end{thebibliography}
\vspace{-30pt}
\end{document}